# Production of Charged Hadrons in Muon Deep Inelastic Scattering

M.S. AL-Buriahi, M. T. Ghoneim and M. T. Hussein

Physics Department, Faculty of Science, Cairo University, 12613 Giza – Egypt

moh_soultan@yahoo.com,  ghoneim@sci.cu.edu.eg   and tarek@Sci.cu.edu.eg

**Abstract**. The production of charged hadrons, in muon Deep inelastic scattering (DIS), at light and heavy target is presented. The particles produced by the interaction with Xenon (Xe) is compared with that produced by the interaction with Deuteron (D), to obtain information on cascading process, forward-backward productions, and the rapidity distribution in different bins of the invariant mass of the interacting system W.

**Keywords**: Rapidity distribution, Particle production sources, Light and heavy targets.

## 1-Introduction

The hadronization process, during Lepton Deep Inelastic Scattering (LDIS), is amongst the most striking phenomena in the particle physics. Its importance is to search for new physics of hadron production. The topic of the production of charged hadrons have been covered by several models [1-4]. Furthermore, the particle production, during muon DIS, has been studied in early experiments, using a free nucleon [5, 6] and bound nucleons (nucleus) targets [7, 8]. Moreover, there are very modern researches [9-12] that study the production of charged hadrons by lepton as projectile. The general features of collision may be understood in the framework of quark-parton model (QPM), in which one or multi photons, emitted by the incident muon, interact with the parton of the target nucleon. In figure 1, this process is described as seen in the center of mass system.

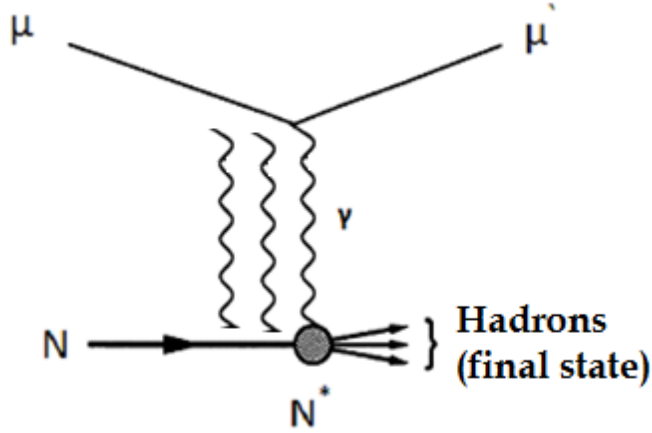

**Figure 1.** Deep inelastic muon-nucleon scattering

The struck parton is emitted into the forward direction, while the remnant of the target travels into the backward direction. Both of these fragments hadronize, to form the forward and backward jets [13].

In our pervious paper [14], we investigated the structure of the nucleon in this process and showed how the hadronic and the leptonic stages. This led us to enrich understanding the inner structure of the nucleon target. In this article, we are going to study the production of the charged hadrons and their correlation in the rapidity space, and in a forthcoming one, we will study the phase transition from QGP to the hadron state in this process.

The production of charged hadrons is discussed in rapidity space. We use Deuterium, as a light nucleus or free nucleon target, and Xenon as a heavy nucleus or bound nucleon target [15]. The results of the two cases will be compared to share some light on the phenomenon of the cascading process.

The paper is organized as follows; after the introduction, we present the rapidity distribution in section 2, the production produced by the interaction with nucleus is compared with that produced by the interaction of nucleon in section 3, and finally the concluding remarks are given in section 4.

## 2- Rapidity Distribution

The lab and cms rapidity are given by

$$y_{lab} = \frac{1}{2} Ln \frac{E_h + p_l}{E_h - p_l} \qquad (1.a)$$

and,
$$y^* = \frac{1}{2} Ln \frac{E_h^* + p_l^*}{E_h^* - p_l^*} \quad (1.b)$$

Where $E_h$, $P_l$ are the energy and longitudinal momentum of the hadron in laboratory frame, the variables in the center-of-mass frame (cms) are labeled by (*). The forward and backward hemispheres correspond to the regions; $y^* > 0$ and $y^* < 0$, respectively.

The difference between the rapidity values in the two frames (Laboratory and center of mass) is constant [16];

$$y^* - y_l = c \quad (2)$$

It is assumed that the particles are produced by different mechanisms/sources. In other words, the overall distribution in a one-collision process is the superposition of particles produced in the central collision (hadronization region), the projectile and target fragmentation.

The analysis of rapidity distribution is performed using the Gaussian parameterization in the form:

$$\frac{dN}{dy} = \sum_i A_i \exp[\frac{-(y-\mu_i)^2}{2\sigma_i^2}] \quad (3)$$

Where $A_i$, $\sigma_i$ and $\mu_i$ are the height, width, and the center of the peak of the successive distributions. Each term may represent a different source (mechanism) of production. Moreover, the area under each Gaussian expresses the weight of each source.

## 2.1. Rapidity Distribution Produced by Muon-Deuteron Collision

The rapidity distributions in different bins of invariant mass, W, for muon-Deuteron collision are shown in figure 2. The fitted distributions are shown as solid lines.

It is found that three Gaussians are sufficient to describe the distributions of low bin values of W. The three Gaussians predicted by the analysis of each curve of figure 2 are displayed in separate graphs. An example is given in figure 3 for the case of particles produced for W=14-20 GeV.

The preliminary analysis predicts that the rapidity distribution of the produced particles is a superposition of two components; namely, the target and projectile fragmentation components at high rapidity that is arising mostly from valence quark-gluon interactions, and the central component at mid-rapidity that is due to gluon-gluon interactions [1].

The parameters $A_i$, $\mu_i$, and $\sigma_i$ (i= 1, 2, 3) of the Gaussians representing the components for each curve are listed in table 1. The area under the Gaussian expresses the weight of each source.

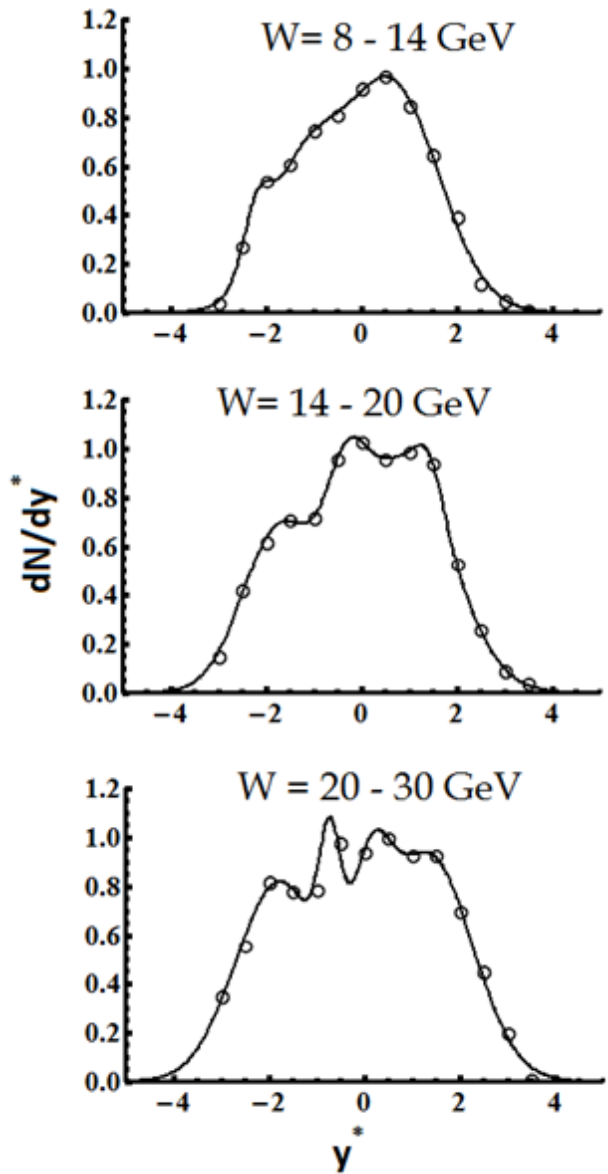

**Figure 2**: Rapidity distributions in different bins of W for Muon-Deuteron Scattering. The lines represent the fitted distributions, assuming three terms of Gaussian shapes.

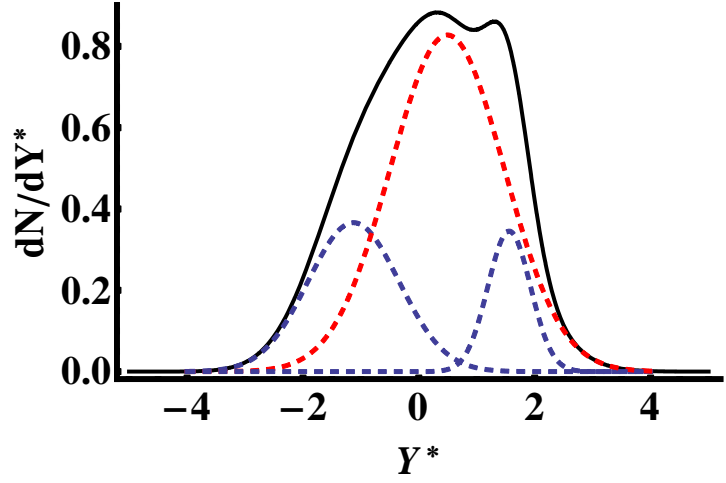

**Figure 3:** Gaussian components of the rapidity distribution (color dashed) in Muon-Deuteron scattering. The red-dashed is the distribution of particles produced in the central rapidity due to gluon–gluon interactions. The blue dashed curves represent the peripheral collisions due to valence quark- gluon. The solid curve is the overall distribution.

**Table 1:** Results of fits of the Gaussian parameterization to the rapidity distribution for charged hadrons in muon-Deuteron scattering, in different bins of W.

| W (GeV) | 8 to 14 | | 14-20 | | 20-30 | |
|---|---|---|---|---|---|---|
| parameters | Values of the parameters | Area | Values of the parameters | Area | Values of the parameters | Area |
| A1 | 0.366 |  | 0.662 |  | 1.077 |  |
| σ1 | 0.615 | 0.72 | 1.004 | 1.661 | 1.049 | 2.762 |
| μ1 | -1.12 |  | -0.96 |  | 0.983 |  |
| A2 | 0.828 |  | 0.243 |  | 0.558 |  |
| σ2 | 0.958 | 2.031 | 0.09 | 0.184 | 0.625 | 1.103 |
| μ2 | 0.491 |  | 2.245 |  | -1.86 |  |
| A3 | 0.345 |  | 0.873 |  | 0.544 |  |
| σ3 | 0.135 | 0.318 | 0.698 | 1.976 | 0.215 | 0.632 |
| μ3 | 1.559 |  | 1.044 |  | -0.64 |  |
| Total Area |  | 3.07 |  | 4.979 |  | 4.497 |

We used the peeling off method to separate the different sources assuming Gaussian form for each. The curves characterized by central peak, with peak near zero in **table 1**, describe the contribution of the gluon–gluon interaction that produces the positive and negative hadrons in the ionization region or non- diffractive process. On the other hand, the other peaks are related to the quark–gluon interaction, producing the charged hadrons due to projectile and target fragmentation or the diffractive processes.

For those of diffractive collisions, the particle going to positive or negative rapidity breaks up. The other incoming particle, still interact with slightly lower momentum, producing particles near the rapidity of the beam.

In a double-diffractive collision, both beam particles break up and produce particles and a dip can be seen in the central region.

It is clear that particle production due to the projectile (muon) fragmentation is substantially less than those in the target fragmentation (nucleons), because of the diversity of their internal structure.

**Table 1** displays the results of the peeling off method. The feature of the peeling off method shows also the ratio of the central area to that of the fragmentation for positive and negative hadrons. In both cases, the increase of the energy W is associated with more valence quark–gluon interaction. This inflates the particle production due to the diffractive process in the fragmentation region, that consequently increases the depth of the dip.

The results displayed in table 2 show that the area of the forward and backward rapidity distributions are equal. This means that the particles are produced in single collision process through the target.

**Table 2:** The percentage of the ration of the production of the particles in backward and forward regions in μ-D collision

| W (GeV) | The ration in backward region (%) | The ration in forward region (%) |
|---|---|---|
| 8 – 14 | 46.1 | 53.8 |
| 14 – 20 | 46.7 | 53.2 |
| 20 – 30 | 50.0 | 49.9 |

## 2.2. Rapidity Distribution Produced by Muon-Xenon Collision

The rapidity distributions in different bins of W for μ -Xe collision are presented in **figure 4**. The same analysis has been done for the μ–Xe interaction as in μ–D. In most cases, four Gaussians were sufficient to fit the experimental data.

The parameters $A_i$, $μ_i$, and $σ_i$ (i= 1→4) of each Gaussian, obtained in the fits of rapidity distribution in μ -Xe collision, are listed in **table 3**

The solid lines in figure 4 represent the superposition of fitted Gaussians. In this case, the area representing the target fragmentation is larger than that for the forward due to the large target size. This leads to a forward-backward symmetry breaking for muon nucleus collision, due to the increase of the cascade collisions in the target nucleus

The Gaussians predicted by the analysis of each case of figure 4 are displayed in separate graphs. An example is given in figure 5 for the case of particles produced at W=14-20 GeV.

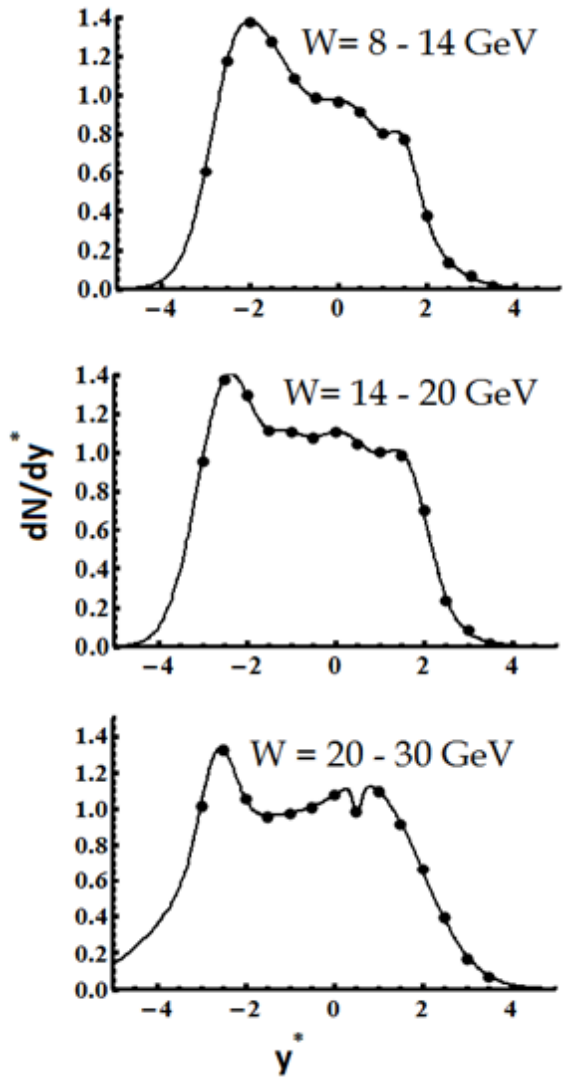

**Figure 4**.Rapidity distributions in different bins of W for Muon-Xenon scattering. The lines represent the superposition of the fitted Gaussian distributions.

**Table 3.** Results of fits of the Gaussian parameterization to the rapidity distribution for charged hadrons in muon – Xenon scattering, in different bins of W.

| W (GeV) | 8 to 14 | | 14-20 | | 20-30 | |
|---|---|---|---|---|---|---|
| parameters | Values of the parameters | Area | Values of the parameters | Area | Values of the parameters | Area |
| A1 | 0.246 |  | 1.1 |  | 0.484 |  |
| σ1 | 0.179 | 0.261 | 1.506 | 3.38 | 0.015 | 0.153 |
| µ1 | -2.51 |  | 0.086 |  | 1.211 |  |
| A2 | 1.098 |  | 0.468 |  | 1.027 |  |
| σ2 | 0.629 | 2.174 | 0.218 | 0.548 | 2.624 | 4.094 |
| µ2 | -1.95 |  | 1.678 |  | -0.528 |  |
| A3 | 0.935 |  | 1.262 |  | 0.872 |  |
| σ3 | 1.409 | 2.779 | 0.462 | 2.128 | 0.271 | 1.133 |
| µ3 | 0.211 |  | -2.452 |  | -2.655 |  |
| A4 | 0.262 |  | 0.249 |  | 0.417 |  |
| σ4 | 0.093 | 0.201 | 0.174 | 0.261 | 0.641 | 0.837 |
| µ4 | 1.527 |  | -1.133 |  | 1.578 |  |
| Total Area | 5.415 | | 6.319 | | 6.217 | |

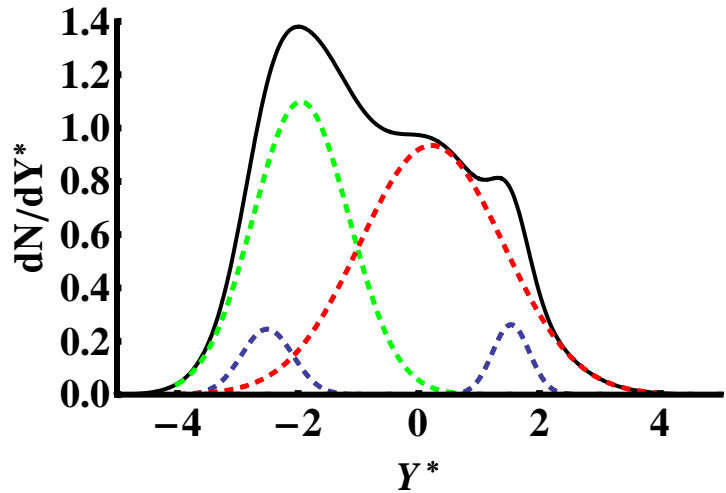

**Figure 5:** Gaussian components of the rapidity distribution (color dashed) in µ -Xenon scattering. Red dashed, represents the central part due to gluon–gluon interaction by non-diffractive process. Blue dashed, represents the peripheral part due to valence quark–gluon interaction through diffractive process. The green dashed represents the excess of particle production due to the multi collision in the target nucleus. The black solid is the super-position that fits the experimental data.

In figure 5, the red Gaussian describes the contribution of the gluon–gluon interaction that produces the charged hadrons in the ionization region or non-diffractive process. The blue Gaussians are related to the quark–gluon interaction that is responsible for producing the charged hadrons due to projectile and target fragmentation or the double diffractive processes of the leading particles. The green Gaussian represents the excess of particles produced due to the multi collisions in the target nucleus.

The average number of collisions inside the target $N_{col}$ is roughly estimated as the ratio between the area of the green Gaussian and that of the blue one seen in the backward region, assuming fixed number of particle production in each collision, which is not exactly true. It is found that $N_{col} = 8$ for the case of Xenon target. In fact, the average number of collisions depends mainly on nucleus size and the collision impact parameter (target thickness). A more refined study for estimating the effective number of collision considering the trajectory of the projectile through the target will be done in details in a forthcoming research.

The backward to forward ratio of particle production is presented in table 4. The excess production in the backward direction is due to the multiple collisions inside the target.

**Table 4:** The percentage of the ration of backward to forward for the particles production in the μ-Xe collision.

| W (GeV) | Total area | The ration in backward region (%) | The ration in forward region (%) |
|---|---|---|---|
| 8 – 14 | 8.5 | 61.9 | 38.0 |
| 14 – 20 | 11.8 | 56.5 | 43.4 |
| 20 – 30 | 12.0 | 58.5 | 41.4 |

## 3- Comparison of Rapidity Distribution Produced by D and Xe

Figure 6 represents a comparison between the theoretical descriptions of rapidity in muon interaction with nucleon and nucleus as targets, at 490 GeV, in three bins of W.

In all cases, the rapidity distribution shows superposition of multiple Gaussians. For muon – Deuteron scattering where single collision is assumed, we used three terms of Gaussian functions to fit the data. However, four terms are necessary to get best fit in the case of muon–Xenon scattering. The Fourth term shows an excess in the production of positive hadrons. This excess depends on the target size.

The forward-backward symmetry is observed in μ-D and breaks for μ-X. The peaks may represent the different sources (mechanisms) of production. The area under the Gaussian expresses the weight of each source. For muon–nucleus, the yield increases rapidly relative to the case of single collision. Moreover, the symmetry breaks for muon nucleus collision because of the increase of the target fragmentation.

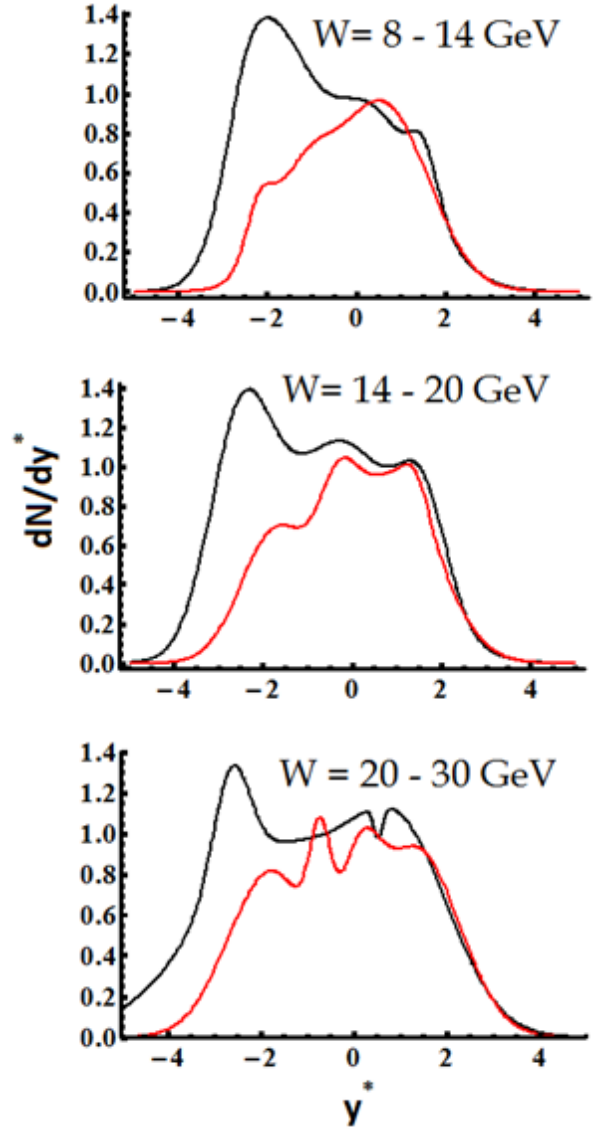

**Figure 6**. Rapidity distributions of positive and negative hadrons for muon – Deuteron (Red line) and muon – Xenon (Black line) interactions, in three bins of W.

## 4- Concluding Remarks

In this work, we studied the production of charged hadrons using nucleon and nucleus as targets in the deep inelastic muon scattering. The Deuteron is used as a free nucleon whichis a good approximation because of; the nucleons in the Deuteron are weakly bound. On the other hand, the Xenon nucleus was a suitable choice to investigate the cascading process; it has a rich numbers of nucleons. We may summarize the results in the following points:

1- In all cases, the rapidity distribution shows superposition of multiple Gaussians. For muon –nucleon scattering where single collision is assumed, we found that three terms of Gaussian function were enough to get fit with the data. However, four terms were found in the case of muon–nucleus scattering. The fourth term shows the excess in the production of hadrons due to multiple collisions in the target nucleus.
2- The rapidity distribution of produced particles is a superposition of two components:
    a. The fragmentation components at high rapidity that are arising mostly from valence quark-gluon interactions producing hadrons through the diffractive process.
    b. The central component at mid-rapidity due to gluon-gluon interactions that produce hadrons through non-diffractive processes.
3- It is clear that particle production due to the projectile (muon) fragmentation is substantially less than those in the target fragmentation (nucleons), because of the diversity of their internal structure.
4- The area under the Gaussians expresses the weight of each source of particle production. It has higher values in the case of muon–nucleus collision. The peeling off method is used to separate the different sources assuming Gaussian form for each.
5- The overall area of the forward and backward production is measured. It has relatively equal values in the case of µ-nucleon scattering, due to the absence of multiple collisions in the target. This symmetry breaks for muon–nucleus collision because of the cascading process in the target.
6- The average number of collisions inside the target Ncol is roughly estimated as the ratio between excess of particle production due to multiple collisions in the target nucleus and that of backward fragmentation region.